\documentclass[apj]{emulateapj} 

\def\cf{cf.}
\newcommand{\be}{\begin{equation}}
\newcommand{\ee}{\end{equation}}

\def\Halpha{\mbox{H\hspace{0.1ex}$\alpha$}}
\def\rmit#1{{\it #1}}
\def\eg{\rmit{e.g.,}}
\def\ie{\rmit{i.e.,}}
\def\edt#1{#1}
 
\def\rmu{{\rm u}}
\def\rml{{\rm l}}

\def\rme{{\rm e}}
\def\rmV{{\rm V}}
\def\rmR{{\rm R}}

\def\exp{\rme}

\def\HI{\ion{H}{1}}

\def\eh2{\ensuremath{e_{\mathrm{H2}}}}
\def\nh2{\ensuremath{n_{\mathrm{H2}}}}
\def\exp{\rme}

\def\Hi{\ion{H}{1}}

\def\h2{\ensuremath{\mathrm{H}_2}}

\def\HI{\ion{H}{1}}
\def\CaII{\ion{Ca}{2}}

\def\MgI{\ion{Mg}{1}}
\def\MgII{\ion{Mg}{2}}
\def\MgIIk{\ion{Mg}{2}\,k}
\def\MgIIh{\ion{Mg}{2}\,h}
\def\MgIII{\ion{Mg}{3}}
\def\hk{{h\&k}}
\def\MgIIhk{\MgII\, \hk}
\def\Bifrost{{\it Bifrost}}
\def\radyn{{\it RADYN}}
\def\multitd{{\it Multi3d}}

\def\RH{{\it RH}}
\def\figspath{.}
\def\lstar{\ensuremath{l^\star}}
\def\ul{\ensuremath{\rmu \rml}}

\def\Pnet{\ensuremath{P_{\mathrm{net}}}}
\def\kthree{\mbox{k$_3$}}    
\def\hthree{\mbox{h$_3$}}
\def\ktwo{\mbox{k$_2$}}
\def\htwo{\mbox{h$_2$}}
\def\kone{\mbox{k$_1$}}     
     
\def\ktwoV{\mbox{k$_{2V}$}}
\def\ktwoR{\mbox{k$_{2R}$}}

\def\ktwov{\mbox{k$_{2V}$}}
\def\ktwor{\mbox{k$_{2R}$}}

\begin{document}

\title{The formation of IRIS diagnostics
  \\ I. A quintessential model atom of \mbox{MG II} and general formation
  properties of the \mbox{MG II H\&K} lines}

   \author{J.~Leenaarts$^{1}$}\email{jorritl@astro.uio.no}
   \author{T. M. D. Pereira$^{2, 3, 1}$}\email{tiago.pereira@astro.uio.no}
   \author{M. Carlsson$^{1}$}\email{mats.carlsson@astro.uio.no}
   \author{H. Uitenbroek$^{4}$}\email{huitenbroek@nso.edu}
   \author{B. De Pontieu$^{3,1}$}\email{bdp@lmsal.com}
 
\affil{$^1$ Institute of
  Theoretical Astrophysics, University of Oslo, P.O. Box 1029
  Blindern, N--0315 Oslo, Norway}
\affil{$^2$NASA Ames Research Center, Mof{}fett Field, CA 94035, USA}
\affil{$^3$Lockheed Martin Solar \& Astrophysics Lab,
         Org.\ A021S, Bldg.\ 252, 3251 Hanover Street
         Palo Alto, CA~94304 USA}
\affil{$^4$NSO/Sacramento Peak P.O. Box 62
         Sunspot, NM 88349--0062 USA}

   \date{Received; accepted}

\begin{abstract}
  NASA's Interface Region Imaging Spectrograph (IRIS) space mission
  will study how the solar atmosphere is energized. IRIS contains an
  imaging spectrograph that covers the \MgIIhk\ lines as well as a
  slit-jaw imager centered at \MgIIk. Understanding the observations
  will require forward modeling of \MgIIhk\ line formation from 3D
  radiation-MHD models. This paper is the first in a series where we
  undertake this forward modeling. We discuss the atomic physics
  pertinent to \hk\ line formation, present a quintessential model
  atom that can be used in radiative transfer computations and discuss
  the effect of partial redistribution (PRD) and 3D radiative transfer
  on the emergent line profiles. We conclude that \MgIIhk\ can be
  modeled accurately with a 4-level plus continuum \MgII\ model atom. Ideally
  radiative transfer computations should be done in 3D including PRD
  effects. In practice this is currently not possible. A reasonable
  compromise is to use 1D PRD computations to model the line profile
  up to and including the central emission peaks, and use 3D transfer
  assuming complete redistribution to model the central depression.
\end{abstract}

  
\section{Introduction}                          \label{sec:introduction}

Magnesium is one of the most abundant elements in the solar atmosphere
and as such its neutral and singly ionized state give rise to a number
of spectral lines with significant diagnostic potential in the photosphere
as well as in the chromosphere. Well-known examples are the \MgI\ 457.1 nm
forbidden line, which is a photospheric temperature diagnostic, the
\MgI\ \textit{b} lines around 517 nm, which serve as upper photospheric/lower
chromospheric magnetic field diagnostics, the \MgI\ 12 $\mu$m infrared lines,
which exhibit very peculiar Non-LTE radiative transfer behavior
\citep{1991ApJ...379L..79C,1992A&A...253..567C},
the 285.2 nm \MgI\ resonance line, which shows effects of partial frequency
redistribution 
\citep[PRD,][]{1995ApJ...447..453U},
 and finally the \MgIIhk\ resonance doublet
at 280.27 and 279.55 nm, respectively. The latter two lines are among
the strongest, and therefore potentially most valuable diagnostically,
in the solar spectrum, but have rarely been observed because of their
wavelength in the middle of the UV range, precluding ground based observation.

This situation will change drastically with the NASA Interface Region
Imaging Spectrograph (IRIS) spacecraft, which will be equipped with a high
resolution UV imaging spectrograph ($<$80\,m\AA\ spectral resolution) and
\MgIIk\ slit-jaw imager (4\,\AA\ filter width) with $0\arcsec.4$
spatial resolution.

Previously, disk center calibrated intensities have been obtained with
the Hawaii
\citep{1978SoPh...60..251A}
and Harvard 
\citep{1976ApJ...205..599K}
sounding rocket experiments. Low temporal and spatial resolution
($2\arcsec$), but high spectral resolution (20\,m\AA)
spectra have been obtained with the
Ultraviolet Spectrometer and Polarimeter
\citep[UVSP,][]{1980SoPh...65...73W} 
instrument on board
the Solar Maximum Mission
\citep[SMM,][]{1984LAstr..98..211V}. 
The lines have been observed
simultaneously with other chromospheric spectral lines in the UV
(Lyman $\alpha$ and  $\beta$, and the \CaII\ H\&K lines) by the 
LPSP instrument on board the \textit{OSO 8} mission
\citep{1978ApJ...221.1032B,1978ApJ...224L..83A,1979SoPh...61...39V,
  1981SoPh...69..289K,1981SoPh...70..325V}
Spatially resolved spectra of the \MgI\ 285.2 nm 
\citep{1995ApJ...447..453U}
 and \MgIIhk\ lines were obtained with the French
balloon experiment RASOLBA
\citep{1995A&A...295..517S}
, and the
sounding rocket experiment HRTS on its ninth flight
\citep{2008ApJ...687..646M},
both on photographic
film. Polarimetric observations in the h\&k lines were obtained with
two flights of the Solar Ultra-violet Magnetograph Investigation
\citep[SUMI,][]{2011SPIE.8160E..29W} 
sounding rocket experiment.

Because magnesium is about 18 times more abundant than calcium
\citep{2009ARA&A..47..481A}
the
h\&k lines form correspondingly higher in the solar atmosphere than
the homologous \CaII\ H\&K lines. From the sparse observations
described above it is clear that the magnesium resonance lines sample a
different regime of the solar atmosphere than the H\&K lines,
which regularly lack emission reversals altogether, or exhibit
singly-peaked profiles
\citep{2008A&A...484..503R},
whereas the former lines always have
doubly-peaked emission reversals, except in sunspots
\citep{2001ApJ...557..854M}.
 In addition, the strong red-blue
asymmetry of the \CaII\ H\&K lines that is thought to be the result
of acoustic waves steepening into shocks
\citep{1997ApJ...481..500C}
is much less pronounced in the h\&k lines
\citep{1989ApJ...337..536G}.
Since the magnetic network gives
rise to more symmetric and less intermittent \CaII\ H\&K line profiles,
the question arises if the magnetic field also gives rise to the
omnipresent emission reversals in the h\&k lines, or whether the
different morphology of the emission in the latter lines can be explained
by the same dynamic effects that cause the intermittent \CaII\ 
behavior, given that the magnesium lines sample a different height
due to their larger opacity.
 
Slight differences in atomic structure between singly ionized calcium
and magnesium create an interesting opportunity for simultaneous
temperature and velocity diagnostics in the spectral window sampled by
the IRIS spectrograph. Whereas the \CaII\ $3d\,^2\!D$ levels have a lower
energy by about 1.3  eV than the $4p\,^2\!P$
upper levels of the H\&K lines, and
thus allow for the $4d\,^2\!D$ -- $4p\,^2\!P$ triplet to form in the
infrared, the \MgII\ $3d\,^2\!D$ levels lie at about the same energy
difference above the $3p\,^2\!P$ levels as these levels lie above the
\MgII\ $3s\,^2\!S$ ground state. As a result the $3d\,^2\!D$ -- $3p\,^2\!P$
triplet forms in the UV, very close to the h\&k resonance lines,
overlapping with them in wavelength.

Because of their high opacity (magnesium is almost exclusively
in its singly ionized state as shown in Sec.~\ref{sec:importance_mgi})
radiation in the
\MgII\ resonance and $3d\,^2\!D$ -- $3p\,^2\!P$ triplet emanates from
in the low-density chromosphere, where non-LTE radiative transfer
and effects of PRD are crucial to line formation.
Previously, the \MgII\ lines had been modeled with PRD in one-dimensional
fashion and compared with observations by
\citet{1974ApJ...192..769M},
 \citet{1997SoPh..172..109U},
and 
\citet{1989ApJ...337..536G},
among others. Inversion of the
spectra that the IRIS mission will provide into physical quantities
is therefore a complicated endeavor and is best achieved by comparing
forward modeling of spectra through numerical simulations of
a radiation magneto-hydrodynamics (RMHD) with observations,
and using such simulations to extract what observable quantities
mean in terms of physical quantities of the underlying atmosphere.
In this paper we therefore present a new updated study of the line formation
properties of the \MgIIhk\ lines incorporating the latest advances in
numerical modeling of the solar chromosphere and non-LTE radiative
transfer to help interpretation of IRIS data. To do so we use the same
snapshot of a radiation-MHD simulation as
\citet{2012ApJ...749..136L},
who studied the formation of the \Halpha\ line. The reader is
encouraged to compare their results with the results for \MgIIhk\ in
the current paper.

In Section~\ref{sec:atomic_data} we describe the atomic data that we
use to construct our \MgII\ model atom. Section~\ref{sec:modelatmos}
describes the model atmospheres we use, Section~\ref{sec:rad_trans}
describes the employed radiative transfer codes. In Section~\ref{sec:noneq}
we investigate the importance of non-equilibrium ionisation of magnesium, in
Section~\ref{sec:atomic_model} we describe the construction of our
quintessential model atom, \ie\ the smallest model atom that
accurately reproduces the \hk\ line, in
Sections~\ref{sec:broadening}--\ref{sec:prd} we describe the basic \hk\ line
formation properties in some detail, and we finish with our discussion
and conclusions in Section~\ref{sec:conclusions}.

\section{Atomic data}            \label{sec:atomic_data}

We use the magnesium abundance of $A_{\mathrm{Mg}} = 7.60 \pm 0.04$ on
the standard logarithmic scale for abundances where
$A_{\mathrm{H}}=12$ from
  \citet{2009ARA&A..47..481A}.
These authors remark that there is no agreement between abundance
determinations from different lines, and the stated abundance is a
straight mean of the abundance determinations for many lines computed
of \MgI\ and \MgII\ in non-LTE using 1D model atmospheres
  \citep{1998A&A...333..219Z,2004MNRAS.350.1127A}.

The atomic energy levels are taken from the NIST database
\citep{nist-db}.
%

The oscillator strengths and photo-ionization cross sections are taken
from from the TOPbase
database\footnote{\url{http://cdsweb.u-strasbg.fr/topbase/topbase.html}}
\citep{1993A&A...275L...5C}. 
Note that some of the photo-ionization cross-sections provided by
TOPbase extend below the ionization threshold energy. These data
points are added for the purpose of average opacity computations, but
should be discarded for non-LTE radiative transfer computations as in
this paper (Mendoza, 2012, private communication).

Excitation by collisions with electrons is treated with data from
  \citet{1995JPhB...28.4879S}.
Collisional excitation and de-excitation between the \hk\ upper
levels by collisions with neutral hydrogen is included according to
  \citet{1988JPhB...21.4165M}.
These authors give tabulated values up to 5000\,K, we use a linear
extrapolation above this temperature.

Auto-ionization, collisional ionization from the ground state by electrons and charge exchange
with protons and neutral hydrogen are done using the recipes by
  \citet{1985A&AS...60..425A}.

Dielectronic recombination is treated
according to
  \citet{1982ApJS...49R.351S}
and collisional ionization from excited states is computed using the
procedure of 
  \citet{1983MNRAS.203.1269B}.

Broadening of the \MgIIhk\ lines by collisions with neutral hydrogen
is treated with the formalism of
\citet{1995MNRAS.276..859A}
using the pertinent coefficients from
\citet{1998MNRAS.300..863B}.

When we discuss the effect of inclusion of \MgI\ in the atomic model
on the emergent intensity from the \MgIIhk\ lines, we use a model atom
including \MgI, \MgII\ and the ground state of \MgIII. This atom is
created by adding the \MgI\ model atom of
\citet{1992A&A...253..567C}
to our \MgII\ model.

\section{Model atmospheres} \label{sec:modelatmos}

In this paper we use a number of different model atmospheres. One is the
semi-empirical model C from
\citet{1993ApJ...406..319F}.
This is a one-dimensional time-independent atmosphere, constructed to
match spatially and temporally averaged observed spectra in the
UV. This atmosphere is used as a starting point in understanding
\hk\ line formation in a simple, static case. This understanding can
then be used in the analysis of more complicated model atmospheres. We
henceforth refer to this model as FALC.

The second model is a snapshot of a 3D RMHD
simulation performed with the \Bifrost\ code
\citep{2011A&A...531A.154G}.
The same snapshot has been used by
\citet{2012ApJ...749..136L}
to investigate \Halpha\ line formation. 

\Bifrost\ solves the equations of resistive MHD on a staggered Cartesian
grid including a variety of physical processes. The simulation
we use here included optically thick radiative transfer in the
photosphere and low chromosphere
\citep{1982A&A...107....1N, 2000ApJ...536..465S, 2010A&A...517A..49H},
parametrized radiative losses in
the upper chromosphere, transition region and corona
\citep{2012A&A...539A..39C},
thermal
conduction along magnetic field lines
\citep{2011A&A...531A.154G}
 and an equation of state that
includes the effects of non-equilibrium ionization of hydrogen
\citep{2007A&A...473..625L}.

The simulation covers a physical extent of $24 \times 24 \times
16.8$\,Mm, with a grid of $504 \times 504 \times 496$ cells, covering
the upper convection zone, photosphere, chromosphere and the lower
corona.  The horizontal grid spacing is 48\,km, the vertical grid
spacing is non-uniform, with a spacing of 19\,km between $z=-1$ and
$z=5$\,Mm, and increasing towards the bottom and top of the
computational domain. The simulation contains a magnetic field with a
an average unsigned strength of 50\,G in the photosphere, concentrated
in the photosphere in two clusters of opposite polarity. We shall
refer to the snapshot from this simulation as 3DMHD.

We also use a snapshot from a 2D simulation performed with
\Bifrost\ by
\citet{2011A&A...530A.124L}.
This simulation was performed with the same physics as 3DMHD, but
had a very weak magnetic field (0.3 G average unsigned flux in the
photosphere). It has a grid size of $325\times512$ points, with a
horizontal grid spacing of 32.5\,km spanning 16.6\,Mm and a
non-uniform vertical grid with a spacing of 28\,km between the bottom
of the computational domain and $z=5$\,Mm, and increasing towards the
top of the computational domain to 150\,km.

Characteristic of this 2D simulation is that it contains large bubbles of
low temperature ($<2000$\,K) gas in the chromosphere, which do not
occur in 3DMHD. At these low temperatures we expect a significant
amount of magnesium in the form of \MgI, so we use the 2D simulation
to extend the range of physical circumstances under which we study
\MgIIhk\ line formation. We call this snapshot 2DMHD.

\section{Radiative transfer} \label{sec:rad_trans}

We perform non-LTE radiative transfer computations with two different
codes. The first is a modified version of \RH\ by
\citet{2001ApJ...557..389U}.
This version was parallelized using MPI so it can efficiently
 solve the radiative transfer in all columns in a 3D model
atmosphere assuming each column is a plane-parallel 1D atmosphere. It
can treat the effects of angle-dependent partial frequency redistribution
(PRD) using the fast approximation by
\citet{2012A&A...543A.109L}.
\edt{We employ a non-constant coherency fraction
  $\gamma$, given by}
\be
\gamma = \frac{P_j}{P_j+Q^{\mathrm{E}}_{j}},
\ee
\edt{where $P_j$ is the total (collisional and radiative) rate out of the
upper level of the transition $j$ and $Q^{\mathrm{E}}_{j}$ is the rate of
elastic collisions for the upper level. For further details on the
treatment of PRD we refer to 
\citet{2001ApJ...557..389U}.
}

The second code we use is \multitd\
\citep{2009ASPC..415...87L}.
This is an MPI-parallelized radiative transfer code that can
evaluate the radiation field in full 3D, taking the horizontal
structure in the model atmosphere into account. The current iterative
solution methods for non-LTE radiative transfer with PRD effects
are not stable enough to include in 3D. Our computations with
\multitd\ are done assuming complete redistribution (CRD). This assumption
gives wrong line intensities from the wings up to and including
the line core emission peaks (\ktwo\ and \htwo). However, we show in
Section~\ref{sec:3d_effects} that the central line depressions (\kthree\
and \hthree) are only weakly affected by PRD. 

\edt{For the FALC model we add additional spectral line broadening using the
  height-dependent microturbulent velocity given in 
\citet{1991ApJ...377..712F}.
Computations using the 3DMHD and 2DMHD atmospheres do not use microturbulence.}

Summarizing, we model the h~and k~lines in 1D including PRD with the
\RH\ code, and in 3D assuming CRD with \multitd. Ideally one should
perform 3D computations including PRD, but this is currently beyond
our capabilities.

%
%


\section{Non-equilibrium ionization} \label{sec:noneq}
Hydrogen ionization is out of equilibrium in the dynamic solar
chromosphere due to long ionization/recombination time-scales 
compared with dynamical timescales
\citep{2002ApJ...572..626C,2007A&A...473..625L}. 
If the same is true for magnesium, it would be necessary
to take into account the history of the atmosphere when interpreting
\MgII\ spectra. 

We here investigate the timescales of ionization/recombination of
\MgII-\MgIII\ using the same methodology as in
\citet{2002ApJ...572..626C}.
We use a time-series of 1D hydrodynamical snapshots computed
with \radyn\
\citep[\eg][]{1992ApJ...397L..59C, 1997ApJ...481..500C}.
This code self-consistently solves the equations of conservation
of mass, momentum, energy and charge together with the non-LTE
rate equations for hydrogen, helium and
calcium. The equations are formulated on an adaptive grid
\citep{1987JCoPh..69..175D} that ensures high numerical resolution
where the gradients are large (\eg\ in shock fronts). We use the same
time-series as in 
\citet{2002ApJ...572..626C} and refer the reader to
that paper for details on the numerical implementation.

\begin{figure}
  \includegraphics[width=\columnwidth]{\figspath/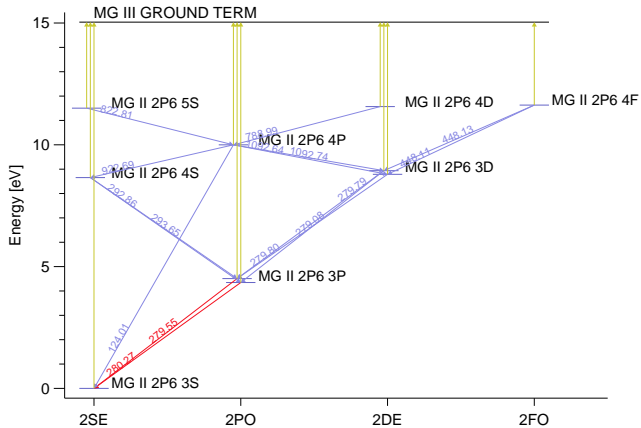}
\caption{Term diagram of the 10-level plus continuum \MgII\ atomic
  model. The \MgIIhk\ lines are indicated by red lines, all other
  bound-bound transitions are blue, the bound-free transitions are
  yellow. All lines indicating bound-bound transitions are annotated
  with the line-center wavelength in nm. \label{fig:11_level_term}}
\end{figure}

For each snapshot in the time-series we perturb the atmosphere by
increasing the temperature by 1\% and calculate the time evolution of
the population densities. We use a 10-level plus continuum model atom
for \MgII. Its term diagram is given in Fig.~\ref{fig:11_level_term}. 
We determine the timescale for ionization/recombination by fitting the time evolution of the \MgIII\
population density with the analytic solution for a two-level atom:

$$n(t)=n(\infty)+(n(0)-n(\infty))\exp^{-t/t_{\rm relax}},$$

where $n(t)$ is the population density at time $t$, $n(\infty)$ is the
equilibrium population density of the perturbed atmosphere, $n(0)$ is
the population density of the initial atmosphere and $t_{\rm relax}$ is the
timescale for relaxation to the final state.

\begin{figure}
  \includegraphics[width=\columnwidth]{\figspath/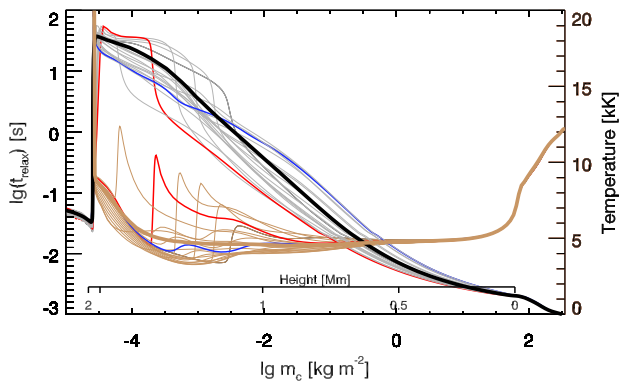}
  \caption{Relaxation timescale (grey/black, left axis) and gas
   temperature (brown, right axis) as function of column mass
   and time in the dynamic simulation at 10s intervals from $t=1600$s
   to $t=1780$s. The thickest lines represent the
   relaxation timescale and the temperature averaged in time over all
   time steps between $t = 1120$s and $t = 3590$s. Two individual time 
   steps are drawn as thick lines: $t = 1640$s (red) and $t = 1760$s (blue). 
   The height scale of the initial state is given as reference.
   \label{fig:noneq}}
\end{figure}

Figure~\ref{fig:noneq} shows the relaxation time and the temperature
as function of column mass. The relaxation time increases from
milliseconds in the photosphere to about 50 seconds at the top of the
chromosphere. Note, however, that this dynamic simulation starts from
a radiative equilibrium atmosphere and it thus has a low temperature
in the chromosphere apart from when shocks travel through. In the
shocks, the relaxation time is lowered by one order of magnitude
compared with the pre-shock, low temperature state (\eg\, at
$t=1760$s when there is a shock at $\lg(m_c)=-3.7$). In the transition
region, the relaxation timescale drops to fractions of a
second. We conclude that whenever the temperature is high enough to give a significant
fraction of \MgIII, the relaxation time is short and statistical
equilbrium is a good approximation for magnesium. 

\section{Quintessential model atom}            \label{sec:atomic_model}

In this section we investigate the effect of the size of the model
atom on the emergent \MgIIhk\ line profiles.

\subsection{Starting atom} \label{subsec:starting_atom}

As a starting atomic model for \MgII\ we use the same 10-level plus continuum
model atom as in Section~\ref{sec:noneq}. Its term diagram is given in
Fig.~\ref{fig:11_level_term}. The \hk\ lines at 279.55\,nm and
280.27\,nm are lines from the $3s\,^2\!S$ ground state to the $3p\,^2\!P$
excited states. There is a triplet of lines between the 
$3p\,^2\!P$ and $3d\,^2\!D$ states
with wavelengths close to the \hk\ lines. One has a wavelength of
279.08\,nm and is located on the blue side of the k core, The other
two are overlapping lines at 279.79\,nm and 279.80\,nm and are located
in between the \hk\ line cores. This triplet can potentially be used
as an upper-photospheric/low-chromospheric velocity
indicator. We plan to investigate the diagnostic value of the triplet
in a future paper. The
4d, 4p and 4f terms are merged into single levels. Their energy
is computed as an average of the energy of the levels in the term,
weighted with their statistical weight.

\subsection{Importance of \MgI} \label{sec:importance_mgi}

\begin{figure*}
  \includegraphics[width=17cm]{\figspath/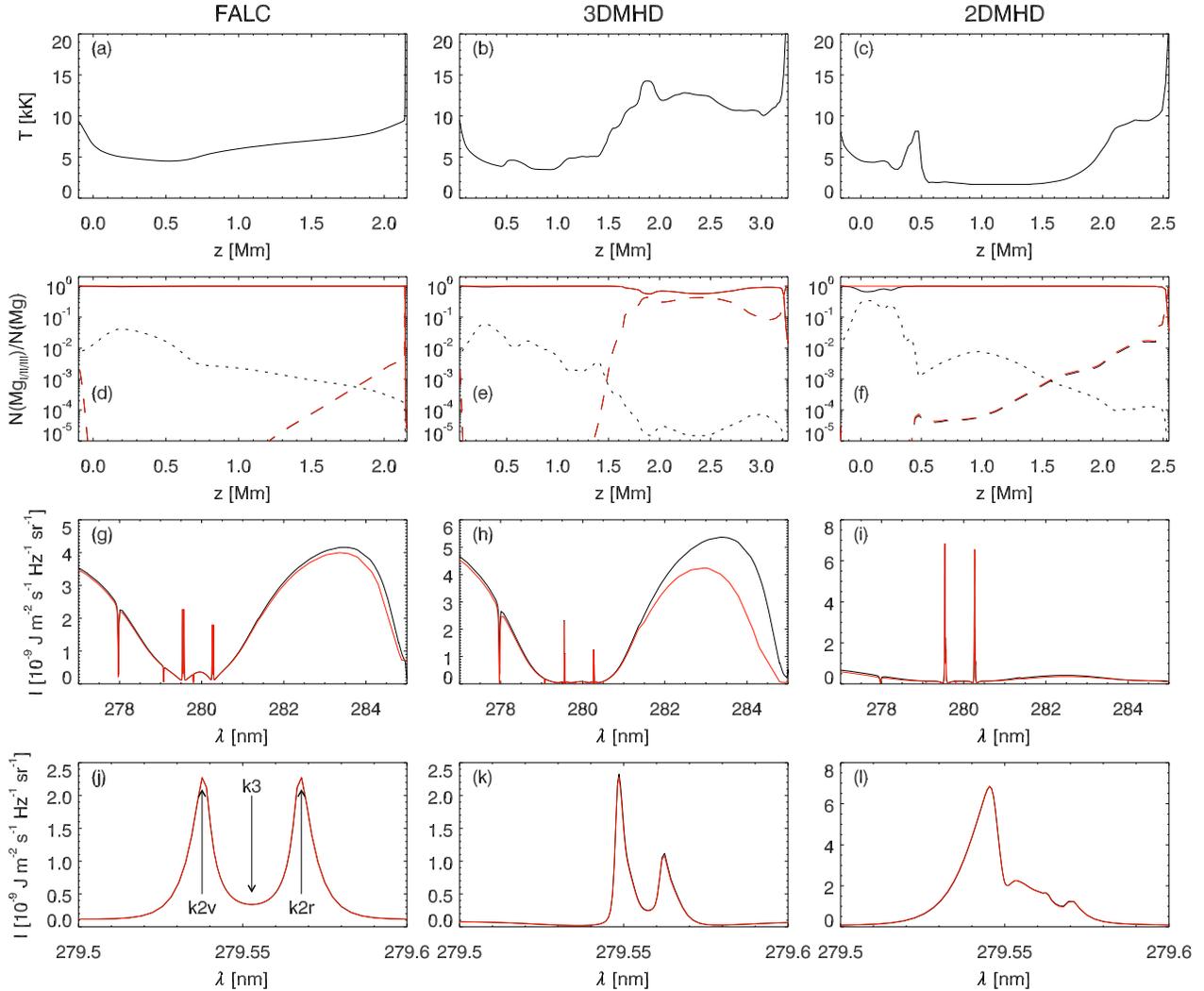}
  \caption{Comparison between the \MgII\ model atoms with 
    \MgI\ treated consistenlty in non-LTE and treated as an LTE source
    of background opacity in three different plane parallel atmospheres (FALC,
    left-hand column, 3DMHD, middle column and 2DMHD, right-hand
    column). Black curves always show quantities for the atom
    including non-LTE \MgI, red curves when \MgI\ is included as
    LTE background opacity.  Panels
    (a)--(c) show the temperature structure as function of height in
    the atmosphere. Panels (d)--(f) display the fraction of all Mg
    atoms in the form of \MgI\ (dotted), \MgII\ (solid) and \MgIII
    (dashed). Dashed: upper
    level of the k line. Dotted: upper level of the h line. Panels
    (g)--(i) show the vertically emergent intensity in the entire
    profile. Panels (j)--(l) show the vertically emergent intensity
    in the core of the k line. In panel (j) we show the commonly used
     names of the
    features in the line core profile. The central depression is called
    \kthree, the emission peak on the short-wavelength (blue) side of 
    the depression is
    called \ktwov, the peak on the red side is called \ktwor.
  \label{fig:MgI_comp}}
\end{figure*}

A fraction of all magnesium atoms in the solar atmosphere is in the
form of \MgI. The ionization balance is sensitive to the temperature
and electron density, through the Saha-equation in LTE and the
ionization and recombination rate coefficients in non-LTE
\citep[\cf][]{1978stat.book.....M,2003rtsa.book.....R}.
It is thus expected that the fraction of magnesium in its neutral
form is highest in cool gas at a high density. 

The addition of \MgI\ levels to our \MgII\ model atom causes a large
performance penalty due to the large increase in the number of levels
and transitions. This penalty is so large that it would
hamper computation of the \hk\ lines from time-series of large 3D
radiation-MHD simulations. 

We therefore investigated whether \MgI\ should be included in non-LTE for
accurate computation of the \hk\ line profiles. We created a model
atom including neutral, singly ionized and twice ionized magnesium by
adding the 65-level \MgI\ atom of
\citet{1992A&A...253..567C}
to the \MgII\ model, including all pertinent transitions.

We then compared the non-LTE radiative transfer problem between
this large model atom, and our small 10-level plus
continuum  \MgII\ atom where we
only included the 65-level \MgI\ atom as a source of background opacity
assuming LTE. Including \MgI\ in LTE in the background does not
significantly increase the amount of computational work.

We used \RH\ to solve the radiative transfer problem for both model atoms in
three plane-parallel atmospheres: FALC, a column of 3DMHD and a column
of 2DMHD. From the latter model we specifically chose a column with a
low temperature throughout most of the chromosphere and photosphere,
thus representing a case with a large influence of \MgI\ on the
\hk\ profiles.

The results are shown in Fig.~\ref{fig:MgI_comp}. Panels (a)--(c) show
the temperature structure in the models, panels (d)--(f) show the
fraction of all magnesium in the various ionization states. As
expected, the fraction of \MgI\ is largest in cool areas in the
photosphere, but only in the extremely cool 2DMHD case it exceeds
10\%. Above 1\,Mm height at most 1\% of magnesium is in the form of
neutral atoms. 

Panels (g)--(o) show the effect on the emergent line profiles. If
\MgI\ is treated as an LTE background element, then the opacity in the
wavelength range of the \hk\ lines is increased owing to the larger
fraction of \MgI\ in LTE compared to non-LTE. In the
\hk\ line cores the \MgI\ opacity is insignificant. In the outer line
wings however, \MgI\ opacity is important, especially in the red wing
of \MgIIh\ because of the presence of the \MgI\ resonance line at
285.2\,nm.

Treating \MgI\ in LTE leads to higher wing opacity. With a source
function that decreases with height this results in a lower \hk\ wing
intensity. The cores of the lines are, however, completely unaffected.


We therefore conclude that it is safe to ignore \MgI\ if one is
interested mainly in the chromosphere where the \hk\ line cores are
formed. If one is interested in precise computation of the outer line
wing intensity, then \MgI\ should be included in non-LTE. Note that the
line wings are blended with many lines, mainly from iron, which
then also should be included. We focus here mainly on the properties
of \hk\ line formation relevant for the IRIS mission, so we ignore
\MgI\ for the remainder of this paper. 

\subsection{Influence of number of levels in the \MgII\ atom}

\begin{figure*}
  \includegraphics[width=\columnwidth]{\figspath/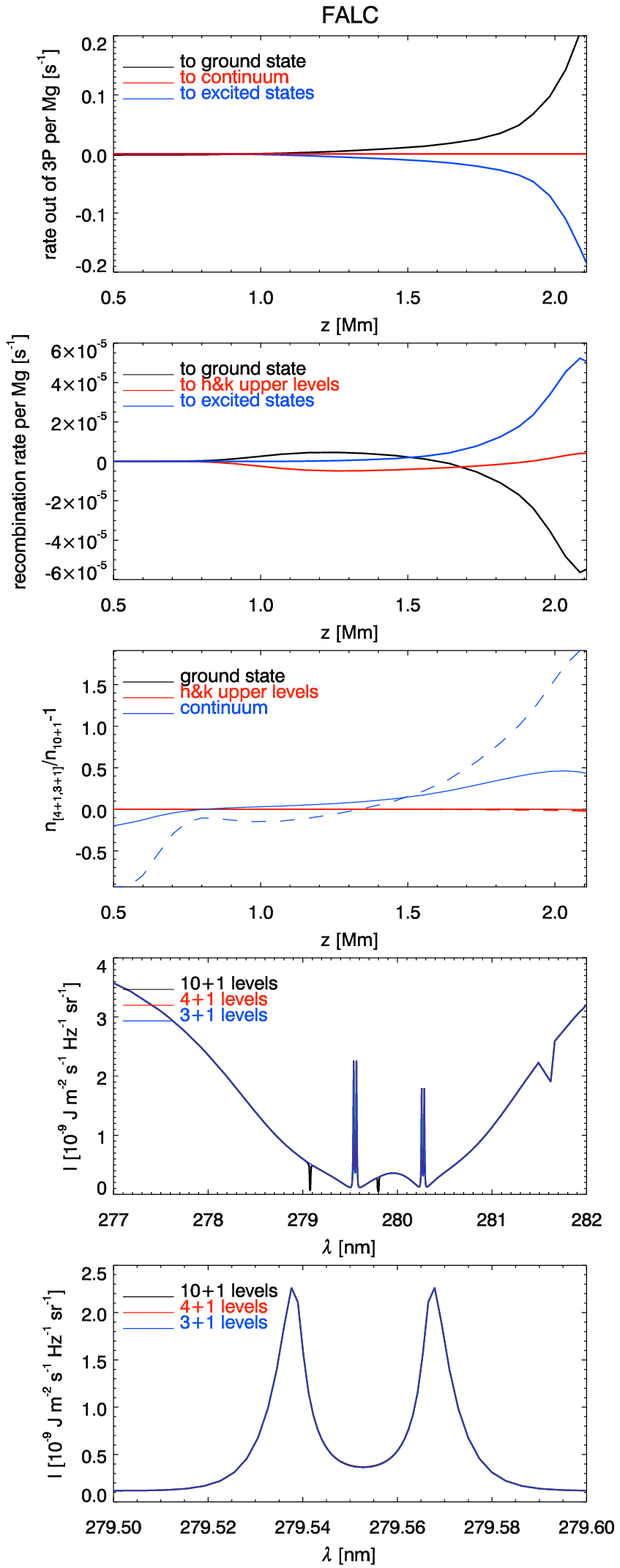}
  \includegraphics[width=\columnwidth]{\figspath/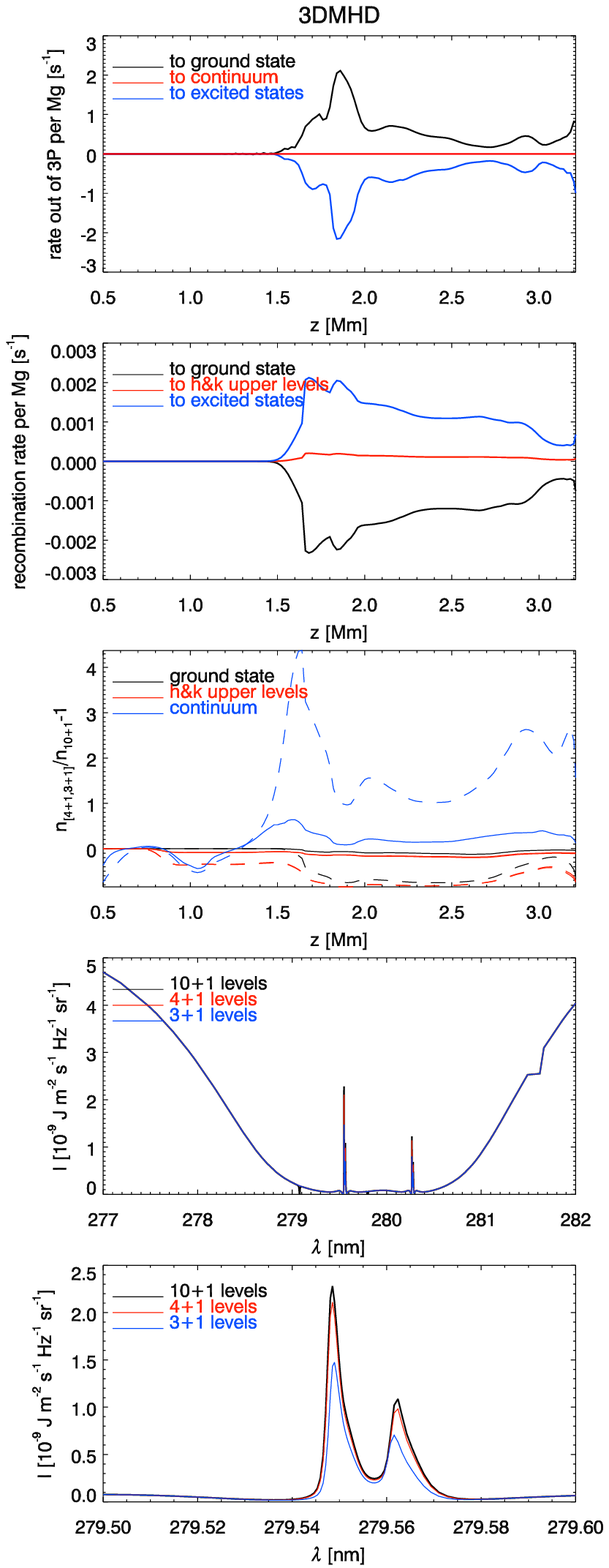}
  \caption{Analysis of the transition rates between levels in the 10+1
    level \MgII\ model atom in the FALC model (left) and a
    column of 3DMHD (right), the same column as in
    Fig.~\ref{fig:MgI_comp}. First row: net transition rate out of
    the \hk\ upper level. Second row: net recombination rate to
    various bound levels. Third row: relative difference in population
    with respect to the 10+1 level reference atom for the 4+1 level
    atom (solid) and the 3+1 level atom (dashed). Fourth row: emergent
    intensity in the \hk lines for the various model atoms. Fifth row:
    emergent intensity in the line core of the k line for the various
    model atoms.
  \label{fig:level_comp}}
\end{figure*}


The 10-level plus continuum \MgII\ atom contains 15 bound-bound and 10
bound-free radiative transitions. Especially for 3D computations it
would be beneficial to construct a model atom with fewer levels
and transitions to speed up computations. To do so we investigated the
various rates in our 10+1 level magnesium atom and show the results of this analysis
in Fig.~\ref{fig:level_comp} for 
the FALC model 
(left-hand column) and a column of 3DMHD (right-hand
column).

The top row shows the net rate $\Pnet$ out of the \hk\ upper levels
($3p\,^2\!P$ states): \be \Pnet = \frac{1}{n_{\mathrm{Mg}}} \sum_i \sum_j \left( n_i
P_{ij} - n_j P_{ji} \right), \ee where the $i$ summation is over the two upper
levels of \hk, and the summation over $j$ is over the target states,
$n_{\mathrm{Mg}}$ is the total density of magnesium atoms, $n_i$ is
the population of level $i$ and $P_{ij}$ is the total (collisional
plus radiative) rate from level $i$ to level $j$. In the deepest part
of the atmosphere the \hk\ lines are in detailed balance, but towards
the top of the atmosphere the \hk\ upper levels are populated from the
higher-lying excited bound states and depopulated to the ground
state. The rate to the continuum is negligible.

The second row shows the net recombination rate to the various
bound levels. Here the situation is more complex. In FALC below
$z=1.5$\,Mm recombination is to the ground state, and ionization is
from the \hk\ upper levels. Above 1.5\,Mm the situation is different,
there ionization is from the ground state, and recombination is to the
higher-lying excited bound states. The same holds for the 3DMHD
column.

In conclusion, the \hk\ lines are effectively in detailed balance from
the photosphere up to the mid chromosphere. In the upper chromosphere,
where the line core forms, \hk\ are part of an ionization/recombination
loop, where recombination occurs through the highly excited levels,
followed by a cascade down through the \hk\ lines to the ground state
after which ionization occurs again.

We then proceeded to study the differences that occur with smaller
model atoms. 

We created the simplest model atom capable of reproducing the
\hk\ lines, which is a 3-level plus continuum atom, with the ground
state and the two $3p\,^2\!P$ states. However, this atom cannot support the
ionization/recombination loop described above.

The smallest model atom that can support the loop is a 4-level plus
continuum atom, the same as the 3+1 level atom, but with an artificial
level included that mimics the effect of the excited states above the
$3p\,^2\!P$ states. This level was created by merging all these excited states
so that the total rates in and out of the levels are roughly conserved
in the new merged level. Its energy is the average of all the merged
levels weighted with their statistical weight.

We then solved the radiative transfer of these 3+1 and 4+1 level atoms
for the same two models, and plot the relative population difference
with the 10+1 level atom in the third row of
Fig.~\ref{fig:level_comp}. In FALC the \hk\ upper and lower
level populations are unchanged, but the population of the continuum
is very different. As expected, the 4+1 level atom does a better job
than the 3+1 level atom in reproducing the full atomic model
populations. In 3DMHD the situation is very different. The 3+1
level atom has too low \hk\ upper and lower level populations, leading
to a too low opacity, and a different  ratio of the upper and lower level
populations, leading to a different source function. The
4+1 level atom does a much better job in reproducing the populations
from the full atom.

This is further demonstrated in the fourth row of
Fig.~\ref{fig:level_comp} which shows the emergent intensity for the
three different model atoms in the \hk\ lines, and the fifth row,
which shows the core of the k line. The line wings are nearly
identical, but the 3+1 level atom fails to reproduce the 10+1 level
atom line core in 3DMHD. In contrast, the 4+1 level atom does
a very good job, with only a slight difference in the \ktwov\ and \ktwor\ peak intensity.

We compared the absolute value of the relative difference in emergent
intensity between the 3+1 and 4+1 level atom and the 10+1 atom for a
$yz$-slice through the 3DMHD snapshot. The 4+1 level atom has a
maximum average deviation of 3\%. Inspection of all individual columns
in the slice shows that the difference in any given column is never
larger than 11\%, whereas the 3+1 level atom shows differences
larger than 50\%

We conclude that it is possible to accurately model the \hk\ lines
with a 4+1 level model atom. Further simplification to a 3+1 level
atom leads to significant errors in the line core intensity.

\section{Broadening in the \hk\ lines} \label{sec:broadening}

\begin{figure}
  \includegraphics[width=\columnwidth]{\figspath/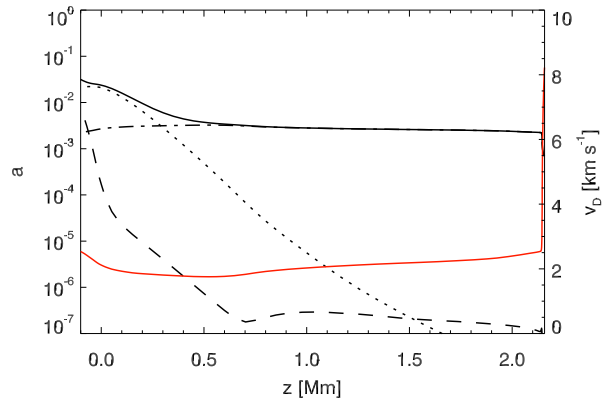}
  \caption{Voigt $a$ damping parameter and Doppler width
    $\sqrt{2kT/m}$ for the \MgIIk\ line in the FALC atmosphere. The
    total damping is shown in solid black (left-hand scale) with the
    contributions from radiative damping (dash-dotted), Van der Waals
    broadening (dotted) and the quadratic Stark effect (dashed). The
    Doppler width is shown in solid red (right-hand scale).
    \label{fig:adamp}}
\end{figure}

In this section we discuss the various broadening mechanisms of the
\hk\ lines. We investigate only the k line, as the h line behaves nearly
identically. Fig.~\ref{fig:adamp} shows the various contributions to
the damping and the Doppler width of the k line in the FALC model.

The \hk\ lines are resonance lines, the radiative damping is the
largest contribution to the total damping in the chromosphere
($z >500$\,km). Van der Waals broadening is the dominant
contribution in the photosphere. Quadratic Stark broadening is only of
minor importance. The Voigt $a$ parameter varies from
$2.5\times10^{-3}$ in the chromosphere to $2.5\times10^{-2}$ in the
photosphere. The \hk\ lines have very broad wings due to their large
opacity in the photosphere. In the FALC model, the combined \hk\ line
opacity at $z=200$\,km becomes equal to the continuum opacity at
277.25\,nm and 282.34\,nm, more than 2\,nm away from the nearest
\hk\ line-center frequency. However, the theoretical continuum
intensity is never reached in a solar spectrum. The haze of
overlapping UV lines prevents the intensity of reaching the `true'
continuum
\citep{1989hsrs.conf..551L}.

Magnesium has an atomic weight of 24.3\,u, and a corresponding Doppler
width $\sqrt{2kT/m}$ between 2 and 3\,km\,s$^{-1}$ for typical
chromospheric temperatures. 

\section{Coupling between \hk\ upper levels and the relative height of
the emission peaks}

\begin{figure}
  \includegraphics[width=\columnwidth]{\figspath/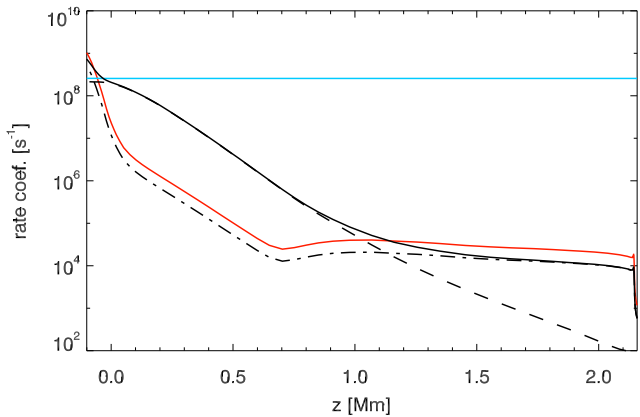}
  \caption{Transition rate coefficients out of the upper level of the
    \MgIIk\ line in the FALC atmosphere. Solid black: total
    collisional rate coefficient to the upper level of the \MgIIh\  line,
    with contribution due to collisions with neutral hydrogen (dashed)
    and electrons (dot-dashed). For comparison the radiative rate
    coefficient to the ground state (blue) and the collisional rate
    coefficient to the ground state (red) are also shown.
    \label{fig:hcol}}
\end{figure}

\begin{figure*}
  \includegraphics[width=\textwidth]{\figspath/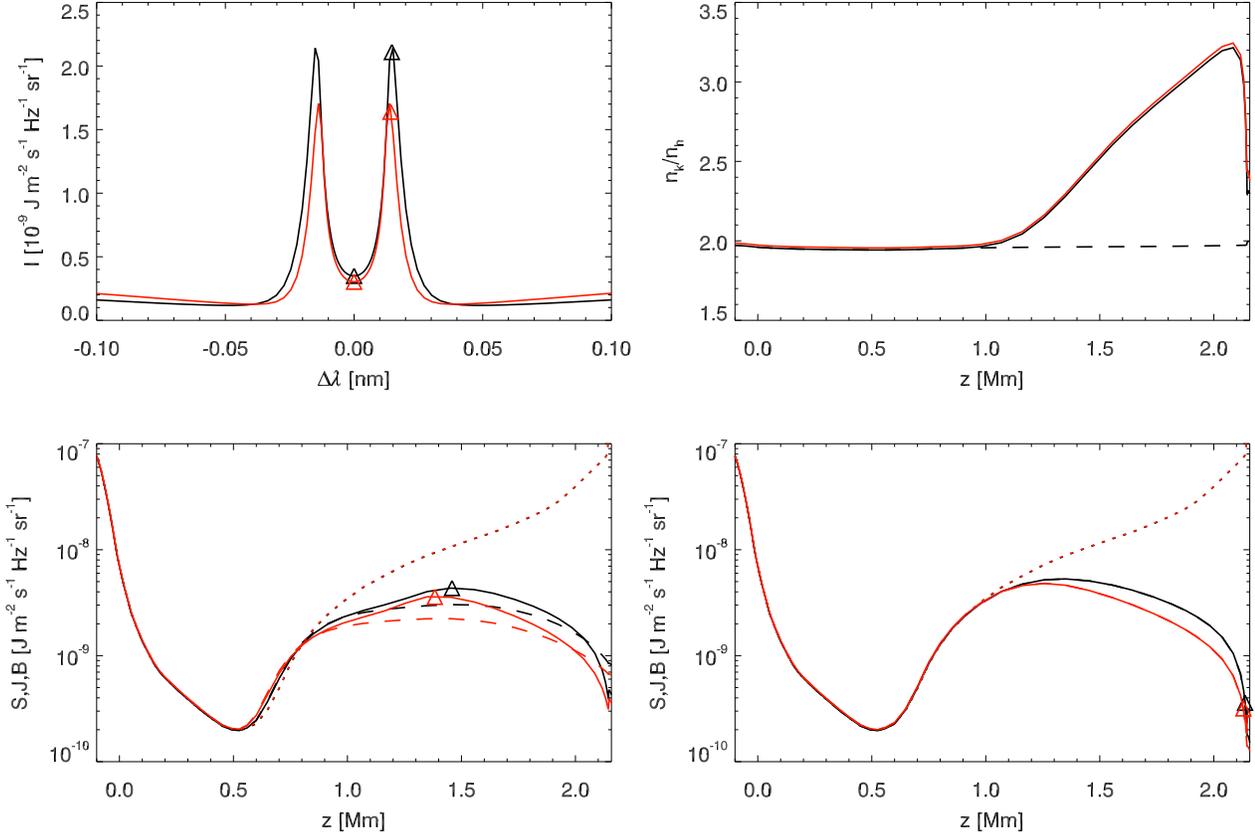}
  \caption{Difference in emission peak height between the \hk\
    lines in the FALC model atmosphere. The triangles indicate
    wavelengths analyzed in the lower panels.  Upper left: h (red) and
    k (black) line core profiles as function of wavelength shift away
    from line center. Upper-right: ratio of the \hk\ upper level
    populations $n_\mathrm{k}/n_\mathrm{h}$ in non-LTE (solid black)
      and LTE (dashed black). The ratio of the upward radiative rates
      $R_\mathrm{k}/R_\mathrm{h}$ is overplotted in red. Lower-left:
        source function (solid), angle-averaged radiation field
        (dashed) and Planck function (dotted) at the red emission peak
        (indicated by the triangles in the upper-left panel) for the
        k line (black) and the h line (red). The triangles indicate
        optical depth unity at this wavelength. Lower-right: same as
        lower left, but now for line-center.
    \label{fig:hkScomp}}
\end{figure*}

Because of the small energy difference between the \hk\ upper levels,
inelastic collisions with neutral hydrogen are important for the
transition rates between these levels, in addition to collisions with
electrons. In the photosphere and lower chromosphere the collisions
with neutral hydrogen are more important than collisions with
electrons due to the low ionization degree of hydrogen there. This is
demonstrated in Fig.~\ref{fig:hcol}, which compares various rate
coefficients out of the upper level of the \MgIIk\ line in the FALC
atmosphere. Collisional de-excitation to the \MgIIh\ upper level by
neutral hydrogen is dominant between 0\,Mm and 1.1\,Mm height, above
1.1\,Mm de-excitation by electrons dominates. For comparison we show
the collisional de-excitation to the ground state (red), which is
comparable to the de-excitation to the \MgIIh\ upper level. In addition
we show the radiative de-excitation to the ground state (blue), which
is orders of magnitude larger than the collisional rates, illustrating
that \MgIIk\ is strongly scattering. The fact that the collisional
rates between the levels are much smaller than the spontaneous
radiative de-excitation rate to the ground state means that the
\hk\ upper levels are only weakly coupled with each other and
significant differences in the source function can be expected. This
is indeed the case, the k line typically has stronger emission peaks
than the h line in both observations
\citep[\eg][]{1967ApJ...149..239T}
and theoretical
calculations. Interestingly the observations also show a larger peak
height ratio in plage than in quiet sun, indication a sensitivity to
the temperature structure in the atmosphere.

The \CaII\,H\&K lines show a similar effect, which was
explained by
\citet{1970SoPh...11..355L}.
In short, the \CaII\,K has twice the opacity as the H line, and hence its source
function thermalizes at a slightly larger height than the source
function of the H line. This means that in a chromosphere with an
outwardly increasing temperature the K source function follows the
temperature rise more strongly than the H line, and hence the K
emission peaks are higher.

\citet{1970PASP...82..169L}
argued that the \MgIIhk\ lines should behave qualitatively the same
as \CaII\,H\&K. We confirm this by analyzing the formation difference
between \hk\ in the FALC atmosphere in
Fig.~\ref{fig:hkScomp}. The upper-left panel shows the \hk\ line
profiles, with a higher emission-peak intensity for the k line. The
upper right panel shows the ratio of the non-LTE upper level populations,
indicating that the k line has a higher source function $S$ because
(neglecting PRD effects) $S_k/S_h \approx {1\over 2}n_k/n_h $ as the lines
share a common lower level and the ratio of the statistical weights of the upper 
levels is 2
\citep[][Sect 2.6.2]{2003rtsa.book.....R}.
The near equality of the population ratio with the ratio of the
radiative rates in the \hk\ lines shows that the populations are set
completely by the line radiation. The lower panels finally show a
comparison of the source function and height of optical depth
unity. Because the k opacity is twice as high as the h opacity,
optical depth unity in the emission peaks in the k line is reached
already at larger heights, and the source function is larger
(lower-left panel). This leads to the larger emission peak height. The
k-line source function is also higher around optical depth unity at line
center, leading to a larger line-center intensity.

\section{Effect of optical pumping by \HI\ Ly\,$\beta$}


The \MgII\ atom has two radiative transitions between the 
$5p\,^2\!P$ states
and the $3s\,^2\!S$ ground state, at wavelengths close to 102.6\,nm, which is
in the red wing of the \Hi\ Ly\,$\beta$ line. Optical pumping in these
transitions from the ground state through the large radiation field
caused by the Lyman line could influence the formation of the
\hk\ lines by creating a pathway for population of the \hk\ upper
levels.

We investigated this effect
by computing the simultaneous non-LTE solution of a five level plus
continuum \HI\ and a 21-level plus continuum \MgII\ atom with and
without these two transitions in the FALC atmosphere and one column
from the 3DMHD snapshot. In both cases neglecting the
$3s\,^2\!S$ -- $5p\,^2\!P$  transitions has a negligible effect
on the \hk\ lines.


\section{Importance of 3D radiative transfer} \label{sec:3d_effects}

\begin{figure}
  \includegraphics[width=\columnwidth]{\figspath/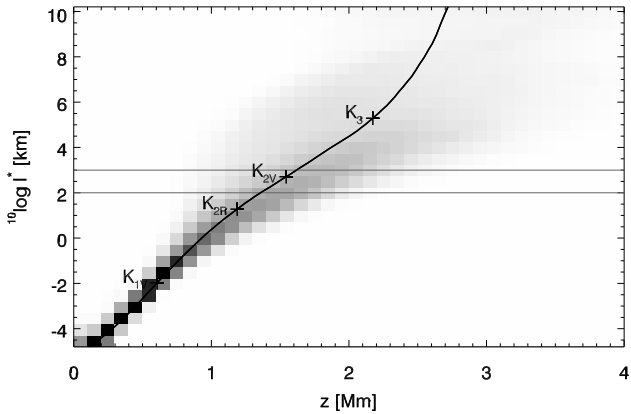}
  \caption{Distribution of 
    photon mean free paths as function of height in the 3DMHD
    model atmosphere at \MgIIk\ line center. The median of the mean
    free path as function of height is shown in solid black. The
    crosses indicate the
    average $\tau=1$ heights at wavelengths for various spectral
    features in the average profile: the minimum on the blue side of
    the central emission peaks (k$_{1 \rmV}$), the red (k$_{2 \rmR}$)
    and blue (k$_{2 \rmV}$)
    emission peaks and the line-center depression (k$_3$). The grey
    lines indicate a free path of 100\,km and 1000\,km to guide the eye.
    \label{fig:lstar}}
\end{figure}

\begin{figure*}
  \includegraphics[width=17cm]{\figspath/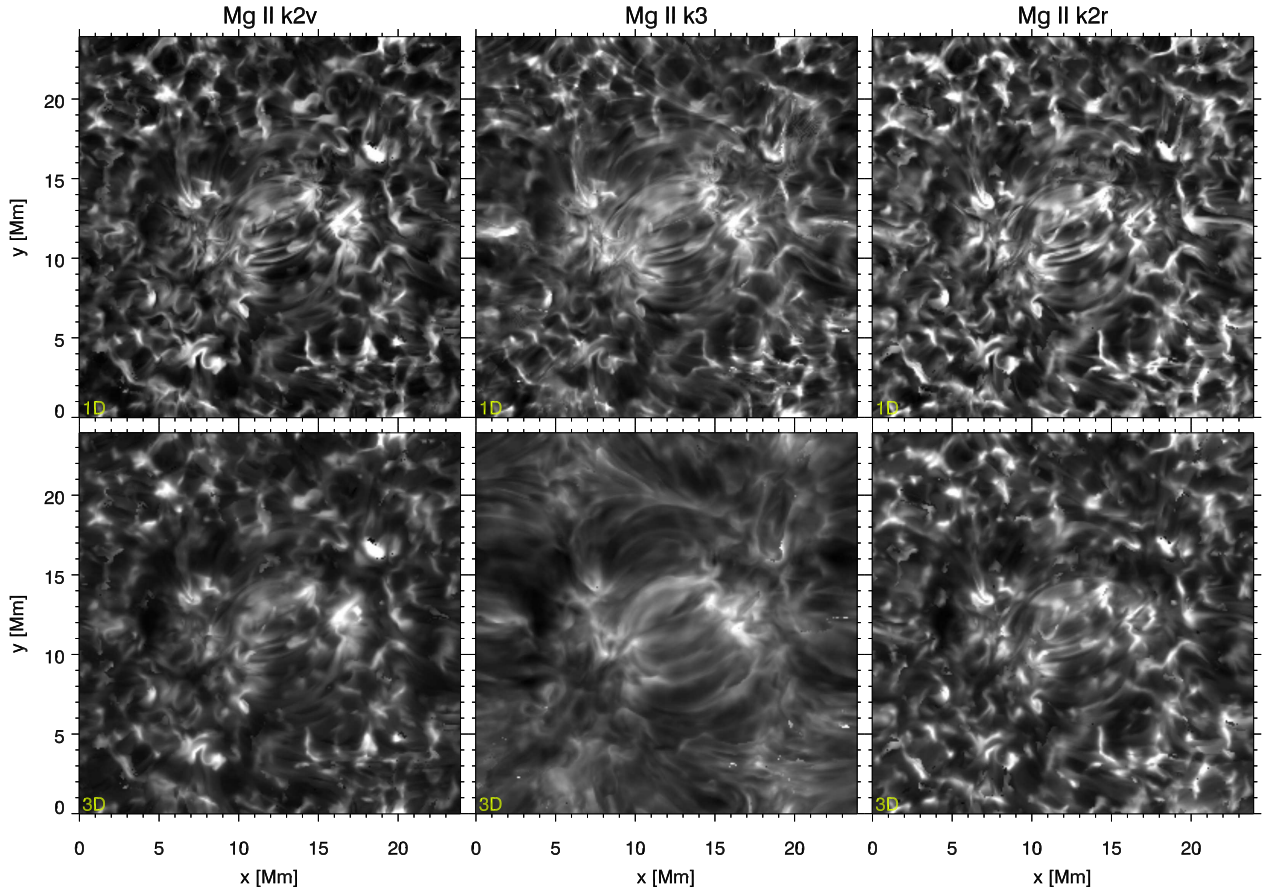}
  \caption{Comparison of the vertically emergent intensity in the
    \MgIIk\ line core between a 1D (top row) and 3D (bottom row)
    computation, performed assuming CRD. The columns show the
    intensity for various spectral features indicated above the top
    panels. The top and bottom panels in each column have the same
    brightness scale so the 1D and 3D intensity can be compared
    directly.
    \label{fig:3d_comp}}
\end{figure*}

\begin{figure*}
  \includegraphics[width=17cm]{\figspath/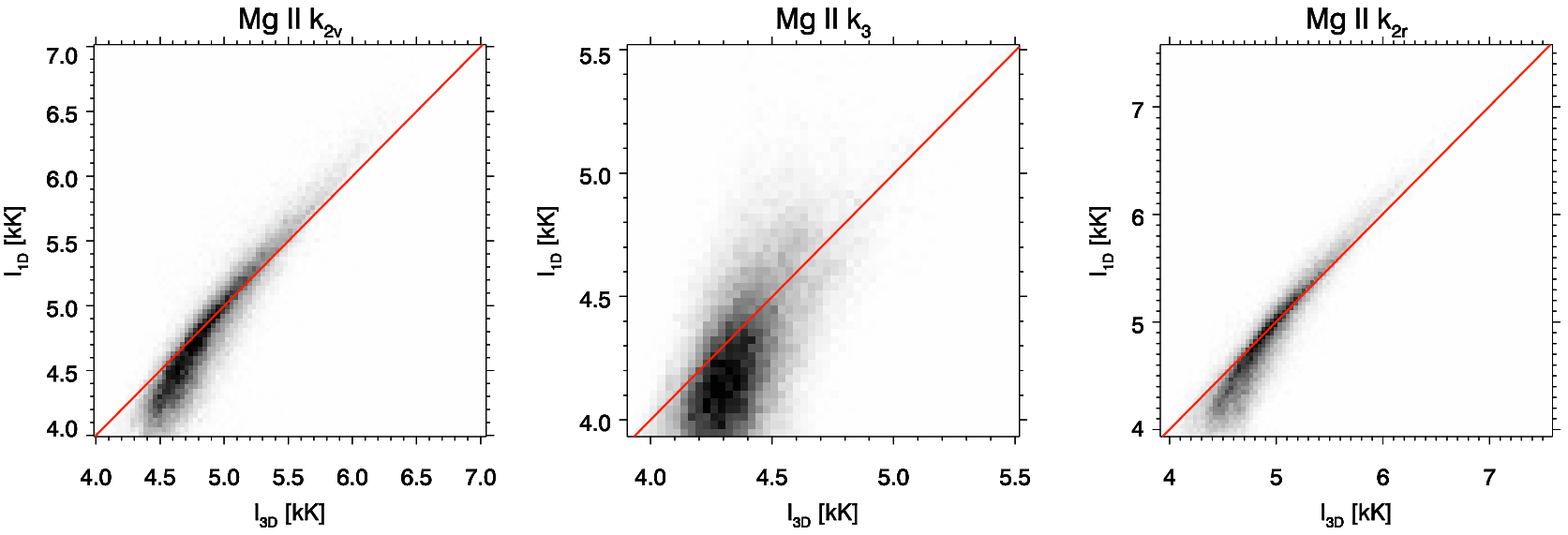}
  \caption{Joint probability density functions of the vertically
    emergent intensity computed in 3D and in 1D. Left: \ktwov; middle:
    \kthree; right: \ktwor. The red line is the line $I_\mathrm{1D}=I_\mathrm{3D}$.
    \label{fig:3d_comp_jpdf}}
\end{figure*}

The \MgIIhk\ lines are strongly scattering as illustrated by
Fig.~\ref{fig:hcol}, which shows that the probability for radiative
de-excitation from the upper levels is 4 orders of magnitude larger
than the probability for collisional de-excitation. This means that
the source function is largely set by the radiation field, which in
turn depends on the atmospheric structure. If the typical effective
photon mean free path is comparable to or larger than the scale of the
horizontal inhomogeneities in the atmosphere then effects of
horizontal radiation transport are important. Typically, this leads to
a smoothing of the radiation field and a corresponding smoothing of
the source function.

We estimate the 3D effects by using the expression for an infinitely
sharp-lined, two-level atom with isotropic scattering.  In that case
the effective photon mean free path can be expressed in terms of the
opacity $\alpha$ and the photon destruction probability $\epsilon$:
\be \label{eq:lstar}
\lstar= \frac{1}{\alpha_\nu \sqrt{\epsilon}}.
\ee
The photon destruction probability is given by
\be \label{eq:epsilon}
\epsilon=\frac{C_{\ul}}{C_{\ul} + A_{\ul} + B_{\ul} B_{\nu}},
\ee
with $C_{\ul}$, $A_{\ul}$ and $B_{\ul}$ the coefficient
for collisional de-excitation, the Einstein coefficients for
spontaneous de-excitation and induced de-excitation, and $B_{\nu}$ the Planck
function.

We computed $\lstar$ for the \MgIIk\ line in the 3DMHD atmosphere,
from a 3D non-LTE CRD computation with \multitd\, taking as opacity
the line-center absorption coefficient for an outgoing vertical
ray. In Fig.~\ref{fig:lstar} we compare it with the average height of
formation of the \kone, \ktwoR, \ktwoV\ and \kthree\ spectral
features. Based on this rough estimate we expect that the \kone\ and
\ktwoR\ intensities are not affected by 3D effects. The \ktwoV\ peak is
marginally affected at the horizontal resolution of the radiative
transfer computation (96\,km, compared to a median photon mean free
path of $\approx 300$\,km), while \kthree\ is strongly affected. 

We confirm this in Figure~\ref{fig:3d_comp}, where we compare the
vertically emergent intensity in \ktwoV, \kthree\ and \ktwoR\ computed
in 3D and 1D, but assuming CRD. For each pixel in the image we
determined the wavelength of the spectral features using an automated
algorithm. Because of the very complex, multiple-peaked profiles, this
algorithm does not correctly identify the spectral locations of the
features for all pixels. These pixel locations show up as noise (dark
pixels in \ktwoV\ and \ktwoR, bright pixels in \kthree). 

The \ktwoV\ and \ktwoR\ spectral features are weakly affected by 3D
effects. Typically, the dark areas are brighter in 3D than in 1D, and
the 3D image is slightly fuzzier. The \kthree\ image is strongly
affected by 3D effects. The 3D image is much fuzzier, and there is a
much more pronounced fibril structure than in 1D. In 1D there are only
fibrils visible in the central region, where the chromospheric
magnetic field is strongest. Elsewhere the 1D image shows shock waves
with a structure very similar to what is seen in the \ktwoV image.

The same information is shown as the joint probability density
function (JPDF) of the 3D and 1D intensities in
Figure~\ref{fig:3d_comp_jpdf}. It confirms the impression from the
images in Figure~\ref{fig:3d_comp}: 3D effects are strong for \kthree,
and much weaker for \ktwov and \ktwor.

For the two emission peaks the largest 3D effects occur at
low intensities, where the 3D intensity is larger than the 1D
intensity. This is caused by the sideways illumination of cool parts
of the atmosphere by hotter neighboring locations. This illumination
leads to a higher angle-averaged radiation field in 3D, and thus to a
higher source function and emergent intensity. The reverse happens for
locations with a high \ktwov and \ktwor\ intensity; there the 3D
radiation field is lower than in 1D owing to the lateral
averaging. The \ktwor\ peak intensity is slightly less influenced by
3D effects than \ktwov, in agreement with the lower formation height
of  \ktwor (see Figure~\ref{fig:lstar}). 

The \kthree\ intensity minimum shows only a weak correlation between
1D and 3D intensity, as expected based on the different appearance of
the 1D and 3D images in Figure~\ref{fig:3d_comp}.

We performed the same analysis for the h line and reached the same conclusions.

\section{Effect of partial redistribution} \label{sec:prd}

\begin{figure}
  \includegraphics[width=8.8cm]{\figspath/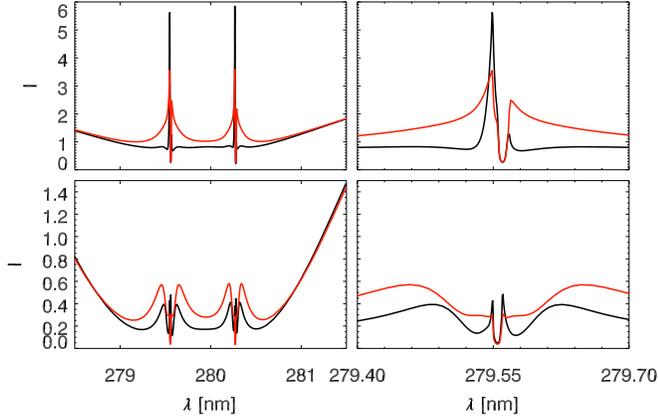}
  \caption{Comparison of the emergent \MgIIhk\ line profiles in two
    locations of the 3DMHD atmosphere (top and bottom
    row). The left-hand column shows both \hk\ lines, the right-hand
    column shows a zoom of the line core of the k line. The black
    curve is the PRD profile, the red curve the CRD profile. The
    intensity is given in units of 10$^{-9}$ J m$^{-2}$ s$^{-1}$ Hz$^{-1}$ sr$^{-1}$.
  \label{fig:prd_crd_comp}}
\end{figure}

\begin{figure*}
  \includegraphics[width=17cm]{\figspath/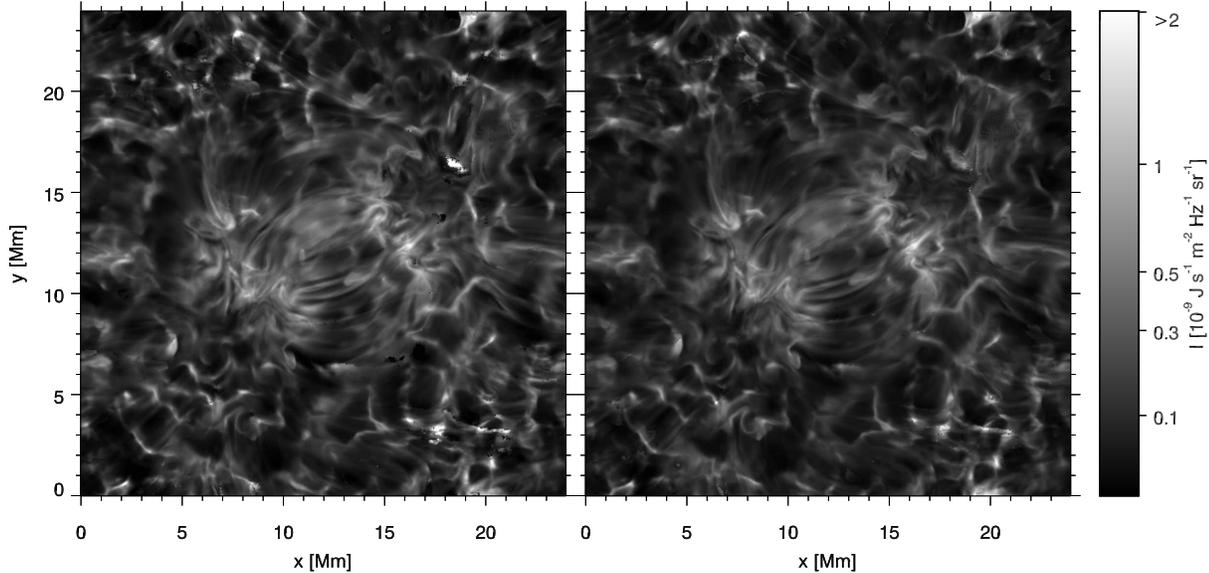}
  \caption{Comparison of the \kthree\ intensity computed in 1D with PRD
    (left-hand panel) and CRD (right-hand panel). Both images have the
    same brightness scale, given on the right. Columns where the wavelength of \kthree\ is
    misidentified appear as white pixels.
  \label{fig:prd_crd_comp_k3_images}}
\end{figure*}

\begin{figure}
  \includegraphics[width=8.8cm]{\figspath/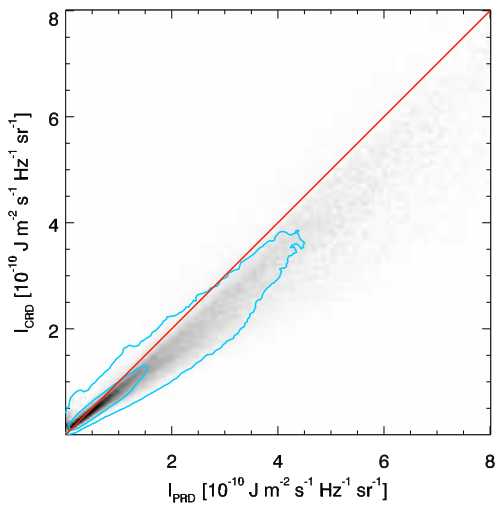}
  \caption{Joint probability density function of the intensity in
    \MgII~\kthree\ computed in PRD (horizontal axis) and CRD (vertical
    axis). Darker color indicates a higher number of pixels in the
    bin. The image has been clipped and scaled to bring out the
    distribution at high intensities. The red line indicates
    $I_\mathrm{PRD}=I_\mathrm{CRD}$. The inner blue contour encloses
    50\% of all pixels, the outer blue contour encloses 90\% of all
    pixels.
  \label{fig:prd_crd_comp_k3_jpdf}}
\end{figure}

We now turn our attention to the effect of PRD on the emergent line
profiles. The \MgIIhk\ lines are strongly affected by the effect of
partial redistribution. Owing to the low particle density in the
chromosphere, the average time between collisions is large compared to
the lifetime of a \MgII\ ion in the upper levels of the
lines. Therefore, the frequency of the absorbed and emitted photon in a
scattering process are correlated and the assumption of complete
redistribution is invalid. We demonstrate this effect in
Fig.~\ref{fig:prd_crd_comp}, which displays a comparison of vertically
emergent line profiles computed with \RH\ in PRD and CRD for two
different columns in the 3DMHD atmosphere. These computations were
done assuming the 1D plane-parallel approximation.

Inclusion of PRD has the strongest effect in the inner wing of the
line profile. Typically the PRD computations yield a lower inner-wing
intensity (around the \kone\ minima) and a higher intensity in the
emission peaks (\ktwoV\ and \ktwoR). The intensity in the
\kthree\ minimum is only weakly influenced by PRD effects. Images of
the \kthree\ intensity computed in PRD and CRD from the 3DMHD
atmosphere appear nearly identical as shown in
Fig.~\ref{fig:prd_crd_comp_k3_images}. 

A more quantitative comparison is given in
Fig.~\ref{fig:prd_crd_comp_k3_jpdf}, which shows the JPDF of the
intensity in \MgII\ \kthree\ computed in PRD and CRD. At low
intensities there is a very tight correlation between the PRD and CRD
computations, at higher intensities the correlation becomes somewhat
weaker and the PRD intensity tends to be higher than the CRD
intensity. In general, assuming CRD for the \kthree\ minimum is a
very good approximation.

%
%

\section{Discussion and conclusions}            \label{sec:conclusions}

We studied the basic formation properties of the
\MgIIhk\ lines in models of the solar atmosphere. 

We first investigated the time dependent, non-equilibrium ionization
balance of \MgII/\MgIII\ and found that whenever the temperature is high enough to give a significant
fraction of \MgIII, the relaxation time is short. We therefore conclude that statistical
equilbrium is a good approximation for the formation of the \MgIIhk\ lines.


We then investigated the influence of \MgI\ on the line formation and
found that treating neutral magnesium in LTE as a source of background
opacity leads to lower intensity in
the outer line wings. The \hk\ line cores are unaffected. If one
is interested in the outer wing intensities, one should include \MgI\ in the
non-LTE model atom. We used a model with 65 bound levels of \MgI. It is
possible that a smaller number of \MgI\ levels is sufficient to model
the effect on the \hk\ lines. If so, this would reduce the required
computational effort significantly. 

We then constructed a minimal model atom that models the \hk\ lines.
This model atom contains four bound levels and a
continuum. The bound levels are the \MgII\ ground state, the 2p$^6$3p
doublet which are the upper levels of the lines, and a higher-lying
artificial level that is required to correctly recover the
\MgII--\MgIII\ ionization balance. This aritificial level acts as a
target level for recombination of \MgIII. Omitting this level leads to
an incorrect ionization balance in the upper chromosphere and too low
emission peaks.

Our investigation of the difference between the h line and the k line
confirm the conjecture by
\citet{1970PASP...82..169L},
that the stronger emission in the k line is because the k line
has twice the opacity of the h line. This larger opacity causes a
larger formation height, and a stronger coupling of the source
function to the gas temperature at the locations in the atmosphere
where the emission peaks form.

The \MgIIhk\ lines are scattering lines, and the emergent intensity is
influenced by 3D effects in the line core. The intensity in the emission peaks
are only mildly affected: in 3D contrast is decreased
compared to 1D computations, but the overall appearance of the images
computed from the 3DMHD model remains the same. The \kthree\ and
\hthree\ intensity minima are strongly affected by 3D effects: 3D
images are fuzzier and show long extended fibrils. Intensity minimum images computed in 1D show only a very weak imprint of fibrils and instead exhibit a shock wave pattern similar to images of the intensity in the emission peaks.

The assumption of complete redistribution is not valid for
\MgIIhk. The line profiles must be computed assuming partial
redistribution, otherwise the line-wing intensity up to and including the emission peaks will be wrong. However, the intensity in the \kthree\ and
\hthree\ minima computed in CRD is very similar to the correct
intensity computed in PRD.

Based on this analysis we can now answer the question how best to
model the \MgIIhk\ lines in time-series of snapshots from large 3D
radiation-MHD models: 

The current state-of-the-art in unpolarized
non-LTE radiative transfer modeling allows for quick computation of
line profiles including PRD treating each column in the snapshot as
an independent 1D plane-parallel atmosphere. Radiative transfer
computations taking the full 3D structure into account are also
possible, but only assuming CRD.  Non-LTE computations including both PRD and 3D simultaneously are
currently not possible: inclusion of PRD in the non-LTE computations
occasionally leads to instabilities in the Lambda-iteration. In 1D
this would only lead to a non-converged column, but in 3D the whole
computation would fail. This is a problem for modeling the \hk\ lines,
which are influenced both by 3D and PRD effects. Our results show, however,
that by combining 1D PRD and 3D CRD computations one can capture the essentials of
the true profile. The profile up to and including the emission
peaks are well approximated by 1D PRD. The central depression can be
modelled reasonably well with 3D CRD computations. We stress that a 1D
CRD approach is not realistic and should not be used.

While this combined approach is not truly satisfactory, it
nevertheless opens up the possibility of modeling the \MgIIhk\ lines
with enough realism to study their formation in 3D numerical models of
the solar atmosphere and determine their potential as atmospheric
diagnostics. We report on the results of such a study in paper II of
this series.

\begin{acknowledgements}
   JL recognizes support from the Netherlands Organization for
   Scientific Research (NWO).  This research was supported by the
   Research Council of Norway through the grant ``Solar Atmospheric
   Modeling'', from the European Research Council under the European
   Union's Seventh Framework Programme \mbox{(FP7/2007-2013)} / ERC Grant
    agreement no. 291058, and through grants of computing time from the Programme
   for Supercomputing of the Research Council of Norway and computing project s1061 from the High End
   Computing Division of NASA. TMDP was supported by the NASA Postdoctoral
    Program at Ames Research Center (grant NNH06CC03B). BDP acknowledges support
   from NASA grants NNX08AH45G, NNX08BA99G, NNX11AN98G, and NNG09FA40C (IRIS).
 We thank Paul Barklem for pointing out the
   paper on collisions with neutral hydrogen.
\end{acknowledgements}


\begin{thebibliography}{55}
\expandafter\ifx\csname natexlab\endcsname\relax\def\natexlab#1{#1}\fi

\bibitem[{{Abia} \& {Mashonkina}(2004)}]{2004MNRAS.350.1127A}
{Abia}, C. \& {Mashonkina}, L. 2004, \mnras, 350, 1127

\bibitem[{{Allen} \& {McAllister}(1978)}]{1978SoPh...60..251A}
{Allen}, M.~S. \& {McAllister}, H.~C. 1978, \solphys, 60, 251

\bibitem[{{Anstee} \& {O'Mara}(1995)}]{1995MNRAS.276..859A}
{Anstee}, S.~D. \& {O'Mara}, B.~J. 1995, \mnras, 276, 859

\bibitem[{{Arnaud} \& {Rothenflug}(1985)}]{1985A&AS...60..425A}
{Arnaud}, M. \& {Rothenflug}, R. 1985, \aaps, 60, 425

\bibitem[{{Artzner} {et~al.}(1978){Artzner}, {Vial}, {Lemaire}, {Gouttebroze},
  \& {Leibacher}}]{1978ApJ...224L..83A}
{Artzner}, G., {Vial}, J.~C., {Lemaire}, P., {Gouttebroze}, P., \& {Leibacher},
  J. 1978, \apjl, 224, L83

\bibitem[{{Asplund} {et~al.}(2009){Asplund}, {Grevesse}, {Sauval}, \&
  {Scott}}]{2009ARA&A..47..481A}
{Asplund}, M., {Grevesse}, N., {Sauval}, A.~J., \& {Scott}, P. 2009, \araa, 47,
  481

\bibitem[{{Barklem} \& {O'Mara}(1998)}]{1998MNRAS.300..863B}
{Barklem}, P.~S. \& {O'Mara}, B.~J. 1998, \mnras, 300, 863

\bibitem[{{Bonnet} {et~al.}(1978){Bonnet}, {Lemaire}, {Vial}, {Artzner},
  {Gouttebroze}, {Jouchoux}, {Vidal-Madjar}, {Leibacher}, \&
  {Skumanich}}]{1978ApJ...221.1032B}
{Bonnet}, R.~M., {Lemaire}, P., {Vial}, J.~C., {et~al.} 1978, \apj, 221, 1032

\bibitem[{{Burgess} \& {Chidichimo}(1983)}]{1983MNRAS.203.1269B}
{Burgess}, A. \& {Chidichimo}, M.~C. 1983, \mnras, 203, 1269

\bibitem[{{Carlsson} \& {Leenaarts}(2012)}]{2012A&A...539A..39C}
{Carlsson}, M. \& {Leenaarts}, J. 2012, \aap, 539, A39

\bibitem[{{Carlsson} {et~al.}(1992){Carlsson}, {Rutten}, \&
  {Shchukina}}]{1992A&A...253..567C}
{Carlsson}, M., {Rutten}, R.~J., \& {Shchukina}, N.~G. 1992, \aap, 253, 567

\bibitem[{{Carlsson} \& {Stein}(1992)}]{1992ApJ...397L..59C}
{Carlsson}, M. \& {Stein}, R.~F. 1992, \apjl, 397, L59

\bibitem[{{Carlsson} \& {Stein}(1997)}]{1997ApJ...481..500C}
{Carlsson}, M. \& {Stein}, R.~F. 1997, \apj, 481, 500

\bibitem[{{Carlsson} \& {Stein}(2002)}]{2002ApJ...572..626C}
{Carlsson}, M. \& {Stein}, R.~F. 2002, \apj, 572, 626

\bibitem[{{Chang} {et~al.}(1991){Chang}, {Avrett}, {Noyes}, {Loeser}, \&
  {Mauas}}]{1991ApJ...379L..79C}
{Chang}, E.~S., {Avrett}, E.~H., {Noyes}, R.~W., {Loeser}, R., \& {Mauas},
  P.~J. 1991, \apjl, 379, L79

\bibitem[{{Cunto} {et~al.}(1993){Cunto}, {Mendoza}, {Ochsenbein}, \&
  {Zeippen}}]{1993A&A...275L...5C}
{Cunto}, W., {Mendoza}, C., {Ochsenbein}, F., \& {Zeippen}, C.~J. 1993, \aap,
  275, L5

\bibitem[{{Dorfi} \& {Drury}(1987)}]{1987JCoPh..69..175D}
{Dorfi}, E.~A. \& {Drury}, L.~O. 1987, Journal of Computational Physics, 69,
  175

\bibitem[{{Fontenla} {et~al.}(1991){Fontenla}, {Avrett}, \&
  {Loeser}}]{1991ApJ...377..712F}
{Fontenla}, J.~M., {Avrett}, E.~H., \& {Loeser}, R. 1991, \apj, 377, 712

\bibitem[{{Fontenla} {et~al.}(1993){Fontenla}, {Avrett}, \&
  {Loeser}}]{1993ApJ...406..319F}
{Fontenla}, J.~M., {Avrett}, E.~H., \& {Loeser}, R. 1993, \apj, 406, 319

\bibitem[{{Gouttebroze}(1989)}]{1989ApJ...337..536G}
{Gouttebroze}, P. 1989, \apj, 337, 536

\bibitem[{{Gudiksen} {et~al.}(2011){Gudiksen}, {Carlsson}, {Hansteen}, {Hayek},
  {Leenaarts}, \& {Mart{\'{\i}}nez-Sykora}}]{2011A&A...531A.154G}
{Gudiksen}, B.~V., {Carlsson}, M., {Hansteen}, V.~H., {et~al.} 2011, \aap, 531,
  A154

\bibitem[{{Hayek} {et~al.}(2010){Hayek}, {Asplund}, {Carlsson}, {Trampedach},
  {Collet}, {Gudiksen}, {Hansteen}, \& {Leenaarts}}]{2010A&A...517A..49H}
{Hayek}, W., {Asplund}, M., {Carlsson}, M., {et~al.} 2010, \aap, 517, A49

\bibitem[{{Kneer} {et~al.}(1981){Kneer}, {Mattig}, {Scharmer}, {Wyller},
  {Artzner}, {Lemaire}, \& {Vial}}]{1981SoPh...69..289K}
{Kneer}, F., {Mattig}, W., {Scharmer}, G., {et~al.} 1981, \solphys, 69, 289

\bibitem[{{Kohl} \& {Parkinson}(1976)}]{1976ApJ...205..599K}
{Kohl}, J.~L. \& {Parkinson}, W.~H. 1976, \apj, 205, 599

\bibitem[{{Leenaarts} \& {Carlsson}(2009)}]{2009ASPC..415...87L}
{Leenaarts}, J. \& {Carlsson}, M. 2009, in Astronomical Society of the Pacific
  Conference Series, Vol. 415, Astronomical Society of the Pacific Conference
  Series, ed. {B.~Lites, M.~Cheung, T.~Magara, J.~Mariska, \& K.~Reeves}, 87

\bibitem[{{Leenaarts} {et~al.}(2011){Leenaarts}, {Carlsson}, {Hansteen}, \&
  {Gudiksen}}]{2011A&A...530A.124L}
{Leenaarts}, J., {Carlsson}, M., {Hansteen}, V., \& {Gudiksen}, B.~V. 2011,
  \aap, 530, A124

\bibitem[{{Leenaarts} {et~al.}(2007){Leenaarts}, {Carlsson}, {Hansteen}, \&
  {Rutten}}]{2007A&A...473..625L}
{Leenaarts}, J., {Carlsson}, M., {Hansteen}, V., \& {Rutten}, R.~J. 2007, \aap,
  473, 625

\bibitem[{{Leenaarts} {et~al.}(2012{\natexlab{a}}){Leenaarts}, {Carlsson}, \&
  {Rouppe van der Voort}}]{2012ApJ...749..136L}
{Leenaarts}, J., {Carlsson}, M., \& {Rouppe van der Voort}, L.
  2012{\natexlab{a}}, \apj, 749, 136

\bibitem[{{Leenaarts} {et~al.}(2012{\natexlab{b}}){Leenaarts}, {Pereira}, \&
  {Uitenbroek}}]{2012A&A...543A.109L}
{Leenaarts}, J., {Pereira}, T., \& {Uitenbroek}, H. 2012{\natexlab{b}}, \aap,
  543, A109

\bibitem[{{Lemaire} \& {Samain}(1989)}]{1989hsrs.conf..551L}
{Lemaire}, P. \& {Samain}, D. 1989, in High spatial resolution solar
  observations, 551

\bibitem[{{Linsky}(1970)}]{1970SoPh...11..355L}
{Linsky}, J.~L. 1970, \solphys, 11, 355

\bibitem[{{Linsky} \& {Avrett}(1970)}]{1970PASP...82..169L}
{Linsky}, J.~L. \& {Avrett}, E.~H. 1970, \pasp, 82, 169

\bibitem[{{Mihalas}(1978)}]{1978stat.book.....M}
{Mihalas}, D. 1978, {Stellar atmospheres /2nd edition/} (San Francisco,
  W.~H.~Freeman and Co., 1978.~650 p.)

\bibitem[{{Milkey} \& {Mihalas}(1974)}]{1974ApJ...192..769M}
{Milkey}, R.~W. \& {Mihalas}, D. 1974, \apj, 192, 769

\bibitem[{{Monteiro} {et~al.}(1988){Monteiro}, {Danby}, {Cooper}, {Dickinson},
  \& {Lewis}}]{1988JPhB...21.4165M}
{Monteiro}, T.~S., {Danby}, G., {Cooper}, I.~L., {Dickinson}, A.~S., \&
  {Lewis}, E.~L. 1988, Journal of Physics B Atomic Molecular Physics, 21, 4165

\bibitem[{{Morrill} {et~al.}(2001){Morrill}, {Dere}, \&
  {Korendyke}}]{2001ApJ...557..854M}
{Morrill}, J.~S., {Dere}, K.~P., \& {Korendyke}, C.~M. 2001, \apj, 557, 854

\bibitem[{{Morrill} \& {Korendyke}(2008)}]{2008ApJ...687..646M}
{Morrill}, J.~S. \& {Korendyke}, C.~M. 2008, \apj, 687, 646

\bibitem[{{Nordlund}(1982)}]{1982A&A...107....1N}
{Nordlund}, A. 1982, \aap, 107, 1

\bibitem[{{Ralchenko} {et~al.}(2011){Ralchenko}, {Kramida}, {Reader}, \& {NIST
  ASD Team}}]{nist-db}
{Ralchenko}, Y., {Kramida}, A., {Reader}, J., \& {NIST ASD Team}. 2011, in
  http://physics.nist.gov/asd

\bibitem[{{Rezaei} {et~al.}(2008){Rezaei}, {Bruls}, {Schmidt}, {Beck},
  {Kalkofen}, \& {Schlichenmaier}}]{2008A&A...484..503R}
{Rezaei}, R., {Bruls}, J.~H.~M.~J., {Schmidt}, W., {et~al.} 2008, \aap, 484,
  503

\bibitem[{{Rutten}(2003)}]{2003rtsa.book.....R}
{Rutten}, R.~J. 2003, {Radiative Transfer in Stellar Atmospheres}, ed. {Rutten,
  R.~J.}

\bibitem[{{Shull} \& {van Steenberg}(1982)}]{1982ApJS...49R.351S}
{Shull}, J.~M. \& {van Steenberg}, M. 1982, \apjs, 49, 351

\bibitem[{{Sigut} \& {Pradhan}(1995)}]{1995JPhB...28.4879S}
{Sigut}, T.~A.~A. \& {Pradhan}, A.~K. 1995, Journal of Physics B Atomic
  Molecular Physics, 28, 4879

\bibitem[{{Skartlien}(2000)}]{2000ApJ...536..465S}
{Skartlien}, R. 2000, \apj, 536, 465

\bibitem[{{Staath} \& {Lemaire}(1995)}]{1995A&A...295..517S}
{Staath}, E. \& {Lemaire}, P. 1995, \aap, 295, 517

\bibitem[{{Tousey}(1967)}]{1967ApJ...149..239T}
{Tousey}, R. 1967, \apj, 149, 239

\bibitem[{{Uitenbroek}(1997)}]{1997SoPh..172..109U}
{Uitenbroek}, H. 1997, \solphys, 172, 109

\bibitem[{{Uitenbroek}(2001)}]{2001ApJ...557..389U}
{Uitenbroek}, H. 2001, \apj, 557, 389

\bibitem[{{Uitenbroek} \& {Briand}(1995)}]{1995ApJ...447..453U}
{Uitenbroek}, H. \& {Briand}, C. 1995, \apj, 447, 453

\bibitem[{{Vial}(1984)}]{1984LAstr..98..211V}
{Vial}, J.-C. 1984, L'Astronomie, 98, 211

\bibitem[{{Vial} {et~al.}(1979){Vial}, {Gouttebroze}, {Artzner}, \&
  {Lemaire}}]{1979SoPh...61...39V}
{Vial}, J.~C., {Gouttebroze}, P., {Artzner}, G., \& {Lemaire}, P. 1979,
  \solphys, 61, 39

\bibitem[{{Vial} {et~al.}(1981){Vial}, {Salm-Platzer}, \&
  {Martres}}]{1981SoPh...70..325V}
{Vial}, J.~C., {Salm-Platzer}, J., \& {Martres}, M.~J. 1981, \solphys, 70, 325

\bibitem[{{West} {et~al.}(2011){West}, {Cirtain}, {Kobayashi}, {Davis}, {Gary},
  \& {Adams}}]{2011SPIE.8160E..29W}
{West}, E., {Cirtain}, J., {Kobayashi}, K., {et~al.} 2011, in Society of
  Photo-Optical Instrumentation Engineers (SPIE) Conference Series, Vol. 8160,
  Society of Photo-Optical Instrumentation Engineers (SPIE) Conference Series

\bibitem[{{Woodgate} {et~al.}(1980){Woodgate}, {Brandt}, {Kalet}, {Kenny},
  {Tandberg-Hanssen}, {Bruner}, {Beckers}, {Henze}, {Knox}, \&
  {Hyder}}]{1980SoPh...65...73W}
{Woodgate}, B.~E., {Brandt}, J.~C., {Kalet}, M.~W., {et~al.} 1980, \solphys,
  65, 73

\bibitem[{{Zhao} {et~al.}(1998){Zhao}, {Butler}, \&
  {Gehren}}]{1998A&A...333..219Z}
{Zhao}, G., {Butler}, K., \& {Gehren}, T. 1998, \aap, 333, 219

\end{thebibliography}
\end{document}